\documentclass[conference]{IEEEtran}

\IEEEoverridecommandlockouts

\usepackage{cite}

\usepackage[english]{babel}

\ifCLASSINFOpdf
   \usepackage[pdftex]{graphicx}
\else
   \usepackage[pdftex]{graphicx}
   \DeclareGraphicsExtensions{.eps}
\fi

\usepackage{grffile}
\usepackage{epstopdf}

\usepackage[tight,footnotesize]{subfigure}

\usepackage{fixltx2e}

\usepackage[cmex10]{amsmath}
\usepackage{color}
\usepackage{array}
\begin{document}

\title{Random Access Congestion Control in DVB-RCS2 Interactive Satellite Terminals}

\author{\IEEEauthorblockN{Alessio Meloni
\thanks{A. Meloni gratefully acknowledges Sardinia Regional Government for the financial support of his PhD scholarship (P.O.R. Sardegna F.S.E. 2007-2013 - Axis IV Human Resources, Objective l.3, Line of Activity l.3.1.).}and Maurizio Murroni\thanks{\copyright 2013 IEEE. The IEEE copyright notice applies. DOI: 10.1109/BMSB.2013.6621777}}

\IEEEauthorblockA{DIEE - Department of Electrical and Electronic Engineering\\
University of Cagliari\\
Piazza D'Armi, 09123 Cagliari, Italy\\
Email: \{alessio.meloni\}\{murroni\}@diee.unica.it}
}

\maketitle

\begin{abstract}
The next generation of interactive satellite terminals is going to play a crucial role in the future of DVB standards. As a matter of fact in the current standard, satellite terminals are expected to be interactive thus offering apart from the possibility of logon signalling and control signalling also data transmission in the return channel with satisfying quality. Considering the nature of the traffic from terminals that is by nature bursty and with big periods of inactivity, the use of a Random Access technique could be preferred. In this paper Random Access congestion control in DVB-RCS2 is considered with particular regard to the recently introduced Contention Resolution Diversity Slotted Aloha technique, able to boost the performance compared to Slotted Aloha. The paper analyzes the stability of such a channel with particular emphasis on the design and on limit control procedures that can be applied in order to ensure stability of the channel even in presence of possible instability due to statistical fluctuations.
\end{abstract}

\begin{IEEEkeywords}
Transmission, Simulation, Channel Modeling, Congestion Control
\end{IEEEkeywords}

\IEEEpeerreviewmaketitle

\section{Introduction}

In consumer type of interactive satellite terminals, users generate a large amount of low duty cycle and bursty traffic with frequent periods of inactivity in the return link. Under these operating conditions, the traditionally used Demand Assignment Multiple Access (DAMA) satellite protocol does not perform optimally, since the response time for the transmission of short bursts can be too long. For this reason, in the recently released specification for the next generation of Interactive Satellite Systems (DVB-RCS2) \cite{DVB}, the possibility of sending logon, control and even user traffic using Random Access (RA) in timeslots specified by the Network Control Center (NCC) is provided to Return Channel Satellite Terminals (RCST). In particular, two methods are considered for RA: the first one is the well known Slotted Aloha (SA) \cite{roberts} \cite{abramson}, the second is called Contention Resolution Diversity Slotted Aloha (CRDSA) \cite{CRDSA1} \cite{CRDSA2} \cite{IRSA}. 

SA represents a well established RA technique for satellite networks in which users send their bursts within slots in a distributed manner, i.e. without any central entity coordinating transmission. While this allows to reach an average throughput around $0.36\ [packets/slot]$ despite the possibility of collision among bursts from different users, SA works with very moderate channel load to ensure acceptable delay and loss probability. This gap has been recently filled with the introduction of a new technique named CRDSA that is able to boost the throughput even up to values close to $1$. This technique is based on the transmission of a chosen number of replicas for each burst payload, similarly to what is done in Diversity Slotted Aloha (DSA) \cite{DSA}. Burst copies are randomly placed within a certain number of slots grouped in a so called frame. However, differently from DSA, in CRDSA each burst copy has a pointer to the location of the other replicas of the same burst payload. Therefore, when the frame arrives at the receiver an interference cancellation (IC) process is accomplished. The IC process consists in removing the interfering content of already decoded bursts from the remaining slots in which collision occurred, in order to try restoring the content of bursts that had all their replicas colliding (from which its name). 
In particular \cite{DVB} specifies two possible variants of CRDSA:
\begin{itemize}
\item Constant Replication Ratio CRDSA (CR-CRDSA): using a constant number of replicas for each burst;
\item Variable Replication Ratio CRDSA (VR-CRDSA): using a number of replicas that can differ among packets, since the number of replicas is determined according to a pre-defined probability distribution.
\end{itemize}
Document \cite{DVB} also claims that the applications using the interactive network may rely on network internal contention control mechanisms to avoid sustained excessive packet loss resulting from simultaneous destructive transmissions but the definition of such a mechanism is claimed to be out of scope for the document. For this reason, in this paper guidelines for the design of such a system having simple yet effective retransmission policies aiming at avoiding congestion at the gateway when using RA in the next generation of satellite terminals are presented. To do so, we rely on the analysis carried out in \cite{stab1}, in which a stability model for the case of CRDSA with geometrically distributed retransmissions was presented. 
The analysis carried out in \cite{stab1} is based on the definition of \textit{Equilibrium Contour} and \textit{Channel Load Line}, similarly to what was done in \cite{stab2} for SA. The Equilibrium Contour represents a set of circumstances for which the ongoing communication is in equilibrium, in the sense that the number of newly transmitted packets is equal to the number of packets successfully sent so that the total number of packets "queuing for retransmission" do not change. The Channel Load Line completes the analysis telling which of these equilibrium points are actually of equilibrium for a given scenario, depending on the user population and on the probability that a user has a new packet to transmit. 
Based on this model it is possible to understand where this points of equilibrium for the ongoing communication are with respect to the throughput. For this reason, this paper discusses parameter settings able to ensure that the communication is taking place in a desirable and stable manner. Moreover, similarly to what was done in \cite{dyn_stab} for SA, control limit procedures able to ensure congestion control in DVB-RCS2 when using CRDSA as transmission mode are discussed.

\section{System Overview}\label{systOV}
\begin{figure}[tbh!]
\centering
\includegraphics [width=9 cm] {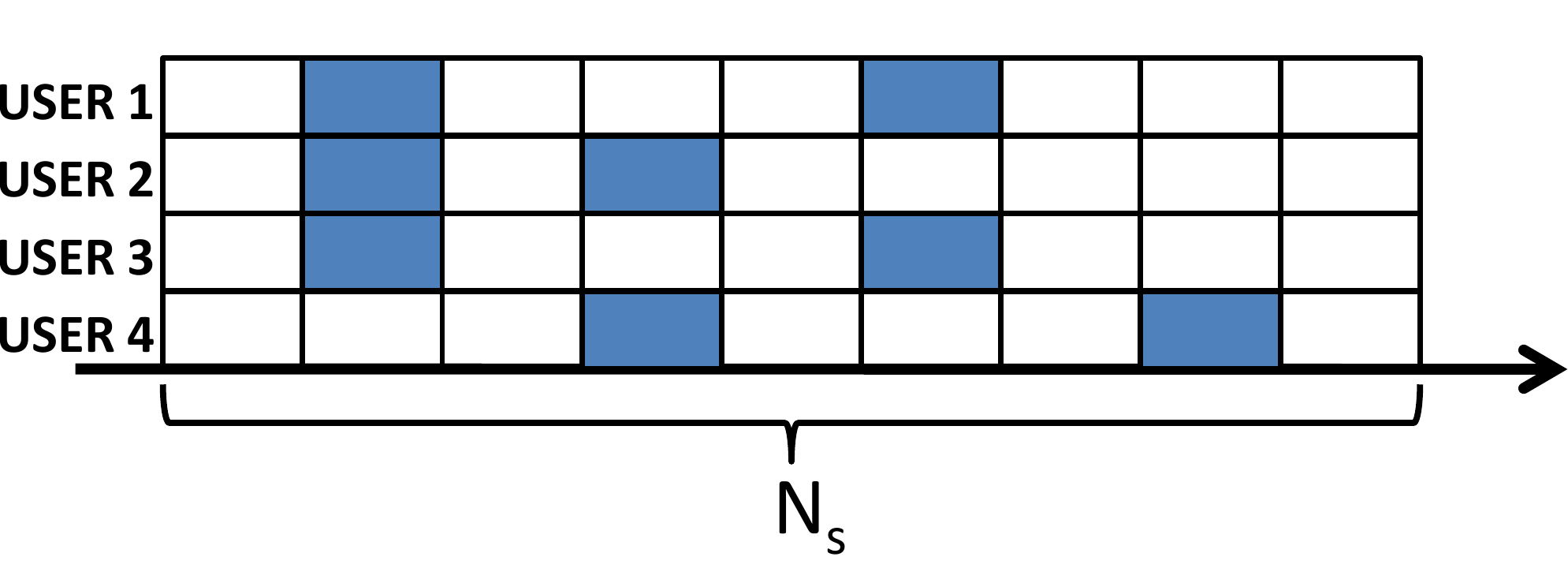}
\caption{Example of frame at the receiver for CRDSA with 2 copies per packet; plain slots indicate that a transmission occurred for that user in that slot.}
\label{Fig1}
\end{figure}

Consider a multi-access channel populated by a total number of users M. Users are synchronized so that the channel is divided into slots and $N_s$ consecutive slots are grouped to constitute a so called frame. The probability that at the beginning of a frame an idle user has a packet to transmit is $p_0$. When a frame starts, users having a packet to transmit place $l$ copies of the same packet over the $N_s$ slots of that frame. The number of copies $l$ can be either the same for each packet or not. Packet copies are nothing else than redundant replicas except for the fact that each one contains a pointer to the location of the others. These pointers are used in order to attempt restoring collided packets at the receiver by means of Successive IC. For the sake of simplicity, throughout the paper perfect IC and channel estimation are assumed which means that the only cause of disturbance for the correct reception of packets is interference among them. Moreover, FEC and possible power unbalance are not considered. Given these conditions, consider the example in Figure~\ref{Fig1} representing a frame at the receiver for the case of 2 copies per packet. Each slot can be in one of three states:

\begin{itemize}
\item{no packet's copies have been placed in a given slot, thus the slot is idle;}
\item{only 1 packet's copy has been placed in a given slot, thus the packet is correctly decoded;}
\item{more than 1 packet's copy has been placed in a given slot, thus resulting in interference of all involved packets.}
\end{itemize}

In DSA, if all copies for a given packet collided the packet is surely lost. In CRDSA, if at least one copy of a certain packet has been correctly received (see User 4), the contribution of the other copies of the same packet can be removed from the other slots. This process might allow to restore the content of packets that had all their copies colliding (see User 2) and iteratively other packets may be correctly decoded up to a point in which no more packets can be restored (or until the maximum number of iterations $I_{max}$ for the IC process is reached, if a limit has been fixed).

When a retransmission process is present, if a packet has not been decoded at the end of the SIC process a feedback is used in order to inform users about the eventual failure of their transmission thus giving birth to a closed loop between the transmitter and the receiver. Since this could cause the channel to get to a state of congestion, a certain policy able to retransmit unsuccessful packets and transmit newly generated packets while keeping the channel in a desirable point of operation is of great interest.

\section{Channel Stability Model}

In this section we quickly revise the channel stability model introduced in \cite{stab1}.
Consider the communication system introduced in Section~\ref{systOV}. Each user can be in one of two states: Thinking (T) or Backlogged (B). Users in T state are idle users that generate a packet for transmission within a frame interval with probability $p_0$; if they do, no other packets are generated until successful transmission for that packet has been acknowledged. Users in B state are users that failed in successfully transmitting their packet and thus are attempting to retransmit it (the probability that they retransmit is $p_r$ in each frame). In the followings, we assume that users are acknowledged about the success of their transmission at the end of the frame (i.e. immediate feedback). From \cite{stab1}, we define

\begin{itemize}
\item{$N_B^f$ : backlogged packets at the end of frame $f$}
\vspace{0.2cm}
\item{$G_B^f=\frac{N_B^{(f-1)}  p_r}{N_s}$ : expected channel load of frame $f$ due to users in B state}
\vspace{0.2cm}
\item{$G_T^f$ : expected channel load of frame $f$ due to users in T state}
\vspace{0.2cm}
\item{$G_{IN}^f=G_T^f+G_B^f$ : expected total channel load of frame $f$}
\vspace{0.2cm}
\item{$PLR^f(G_{IN}^f,N_s,l,I_{max})$ : expected packet loss ratio of frame $f$, with dependence on the expected total channel load $G_{IN}$, the frame size $N_s$, the burst degree distribution $l$ and the maximum number of iterations for the SIC process $I_{max}$}
\vspace{0.2cm}
\item{$G_{OUT}^f=G_{IN}^f ( 1 - PLR^f(G_{IN}^f,N_s,l,I_{max}) )$ : part of load successfully transmitted in frame $f$, i.e. throughput.}
\end{itemize}
\vspace{0.2cm}

 The \textit{equilibrium contour} can then be described as

\begin{equation}\label{gt}
 G_T=G_{OUT}=G_{IN} ( 1 - PLR(G_{IN},N_s,l,I_{max}) )
\end{equation}

\begin{equation}\label{nb2}
N_B=\frac{G_{IN} PLR(G_{IN},N_s,d,I_{max}) N_s}{p_r}
\end{equation}

For the \textit{channel load line}, in the case of finite population 

\begin{equation}\label{LL1}
G_T=\frac{M-N_B}{N_s}p_0
\end{equation}

while for $M\rightarrow\infty$ the channel input can be described as a Poisson process with expected value $\lambda$ [thinking users] so that 
\begin{equation}\label{LL2}
G_T=\frac{\lambda}{N_s}
\end{equation}
 for any $N_B$ (i.e. the expected channel input is constant and independent on the number of backlogged packets). Notice that the frame number $f$ is not present, since in equilibrium conditions always the same value of throughput and backlogged users is expected.

\subsection{Definition of Stability} \label{Stab_Def}

\begin{figure}[tbh!]

\subfigure [Stable channel] {\label{stable} \includegraphics [ scale = 0.22 ]{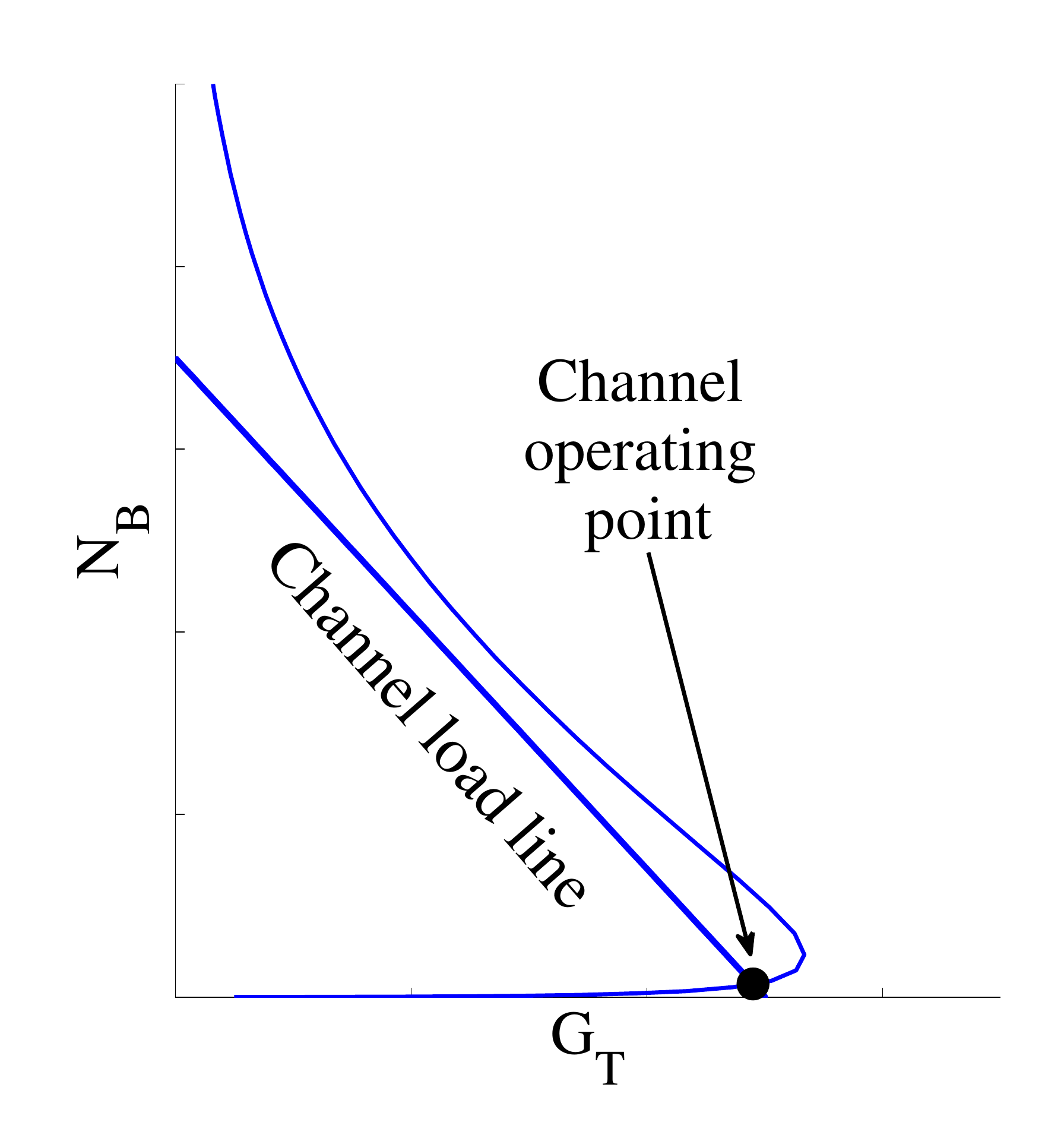}
\label{st}
} \qquad
\subfigure [Unstable channel (finite M)] {\label{unstableFin} \includegraphics [ scale = 0.22 ]{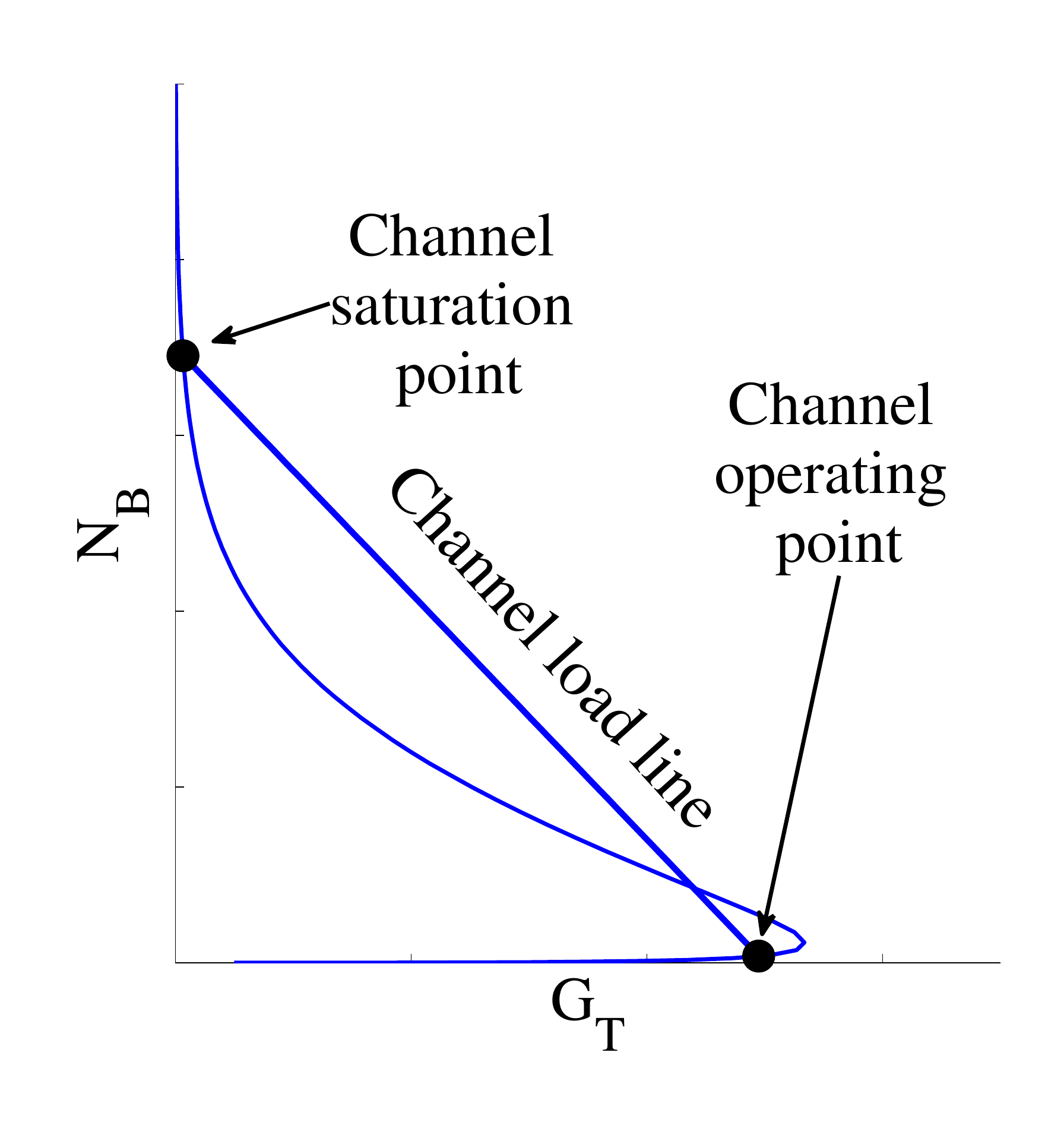}
\label{unstFin}
} \qquad

\subfigure [Unstable channel (infinite M)] {\label{unstableInf} \includegraphics [ scale = 0.22 ]{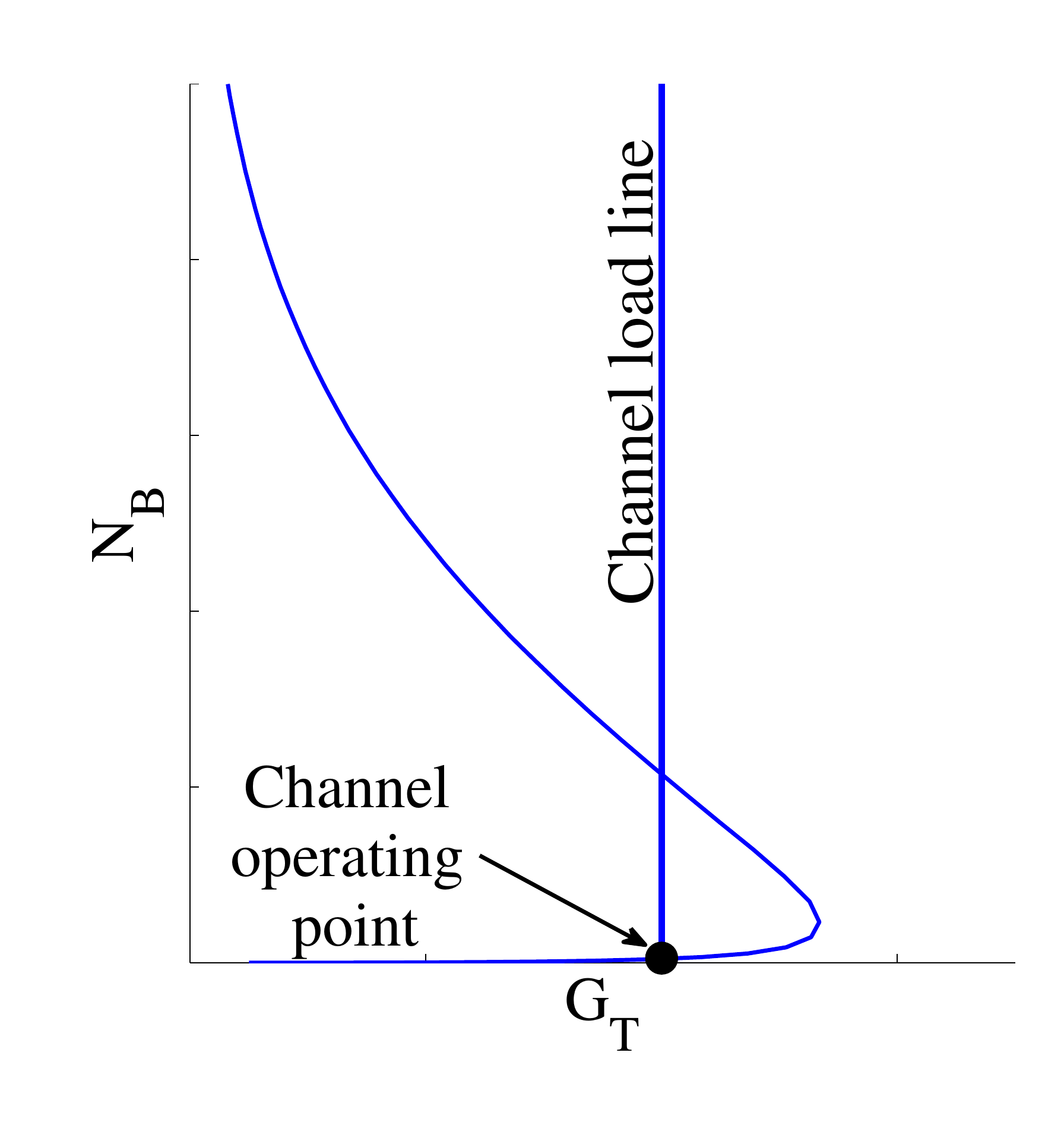}
\label{unstInf}
} \qquad
\subfigure [Overloaded channel] {\label{overL} \includegraphics [ scale = 0.22 ]{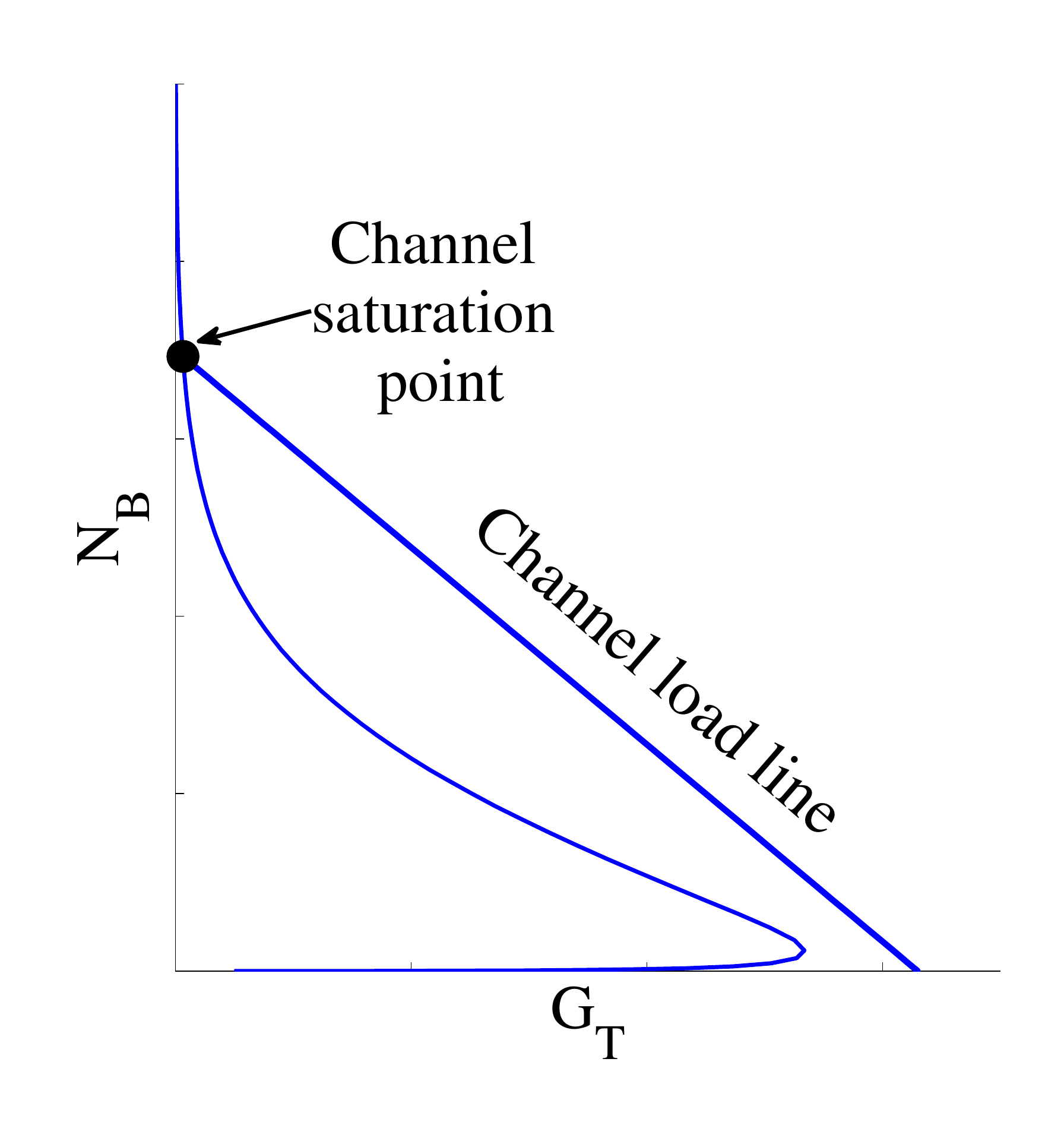}
\label{over}
} \qquad

\caption{\small{Examples of stable and unstable channels}}
\label{All_channels}
\end{figure}

Equilibrium contours represented in Figure~\ref{All_channels} divide the ($N_B$,$G_T$) plane in two parts and each channel load line can have one or more intersections with the equilibrium contour. These intersections are referred to as equilibrium points since $G_{OUT}=G_T$. The rest of the points of the channel load line belong to one of two sets: those on the left of the equilibrium contour represent points for which $G_{OUT}>G_T$, thus situations that yield to decrease of the backlogged population; those on the right represent points for which $G_{OUT}<G_T$, thus situations that yield to growth of the backlogged population.  

From the considerations above, we can gather that an intersection point where the channel load line enters the left part for increasing backlogged population corresponds to a \textit{stable equilibrium point}, since it acts as a sink. In particular, if the intersection is the only one, the point is a \textit{globally stable equilibrium point} (indicated as $G_T^G$,$N_B^G$) while if more than one intersection is present, it is a \textit{locally stable equilibrium point} (indicated as $G_T^S$,$N_B^S$). If an intersection point enters the right part for increasing backlogged population, it is said to be an \textit{unstable equilibrium point} (indicated as $G_T^U$,$N_B^U$) in the sense that as soon as a statistical variation from the equilibrium point occurs, the communication will diverge in one of the two directions of the channel load.

Figure~\ref{st} shows a stable channel. The globally stable equilibrium point can be referred as \textit{channel operating point} in the sense that we expect the channel to operate around that point. With the word around we mean that due to statistical fluctuations, the actual $G_T$ and $N_B$ (and thus also $G_{IN}$ and $G_{OUT}$) may differ from the expected values. In fact, the model deals with the expected values of $G_T$ and $G_B$, and the actual values have binomially distributed probability for $G_B$ and $G_T$ with finite M and Poisson distributed probability for $G_T$ with infinite $M$. Nevertheless, averaging over the entire history of the transmission, simulations in \cite{stab1} have shown that values close to the expected ones are obtained thus validating the assumption of considering the expected value.
 
\begin{figure}[tbh!]
\centering
\includegraphics [scale = 0.35] {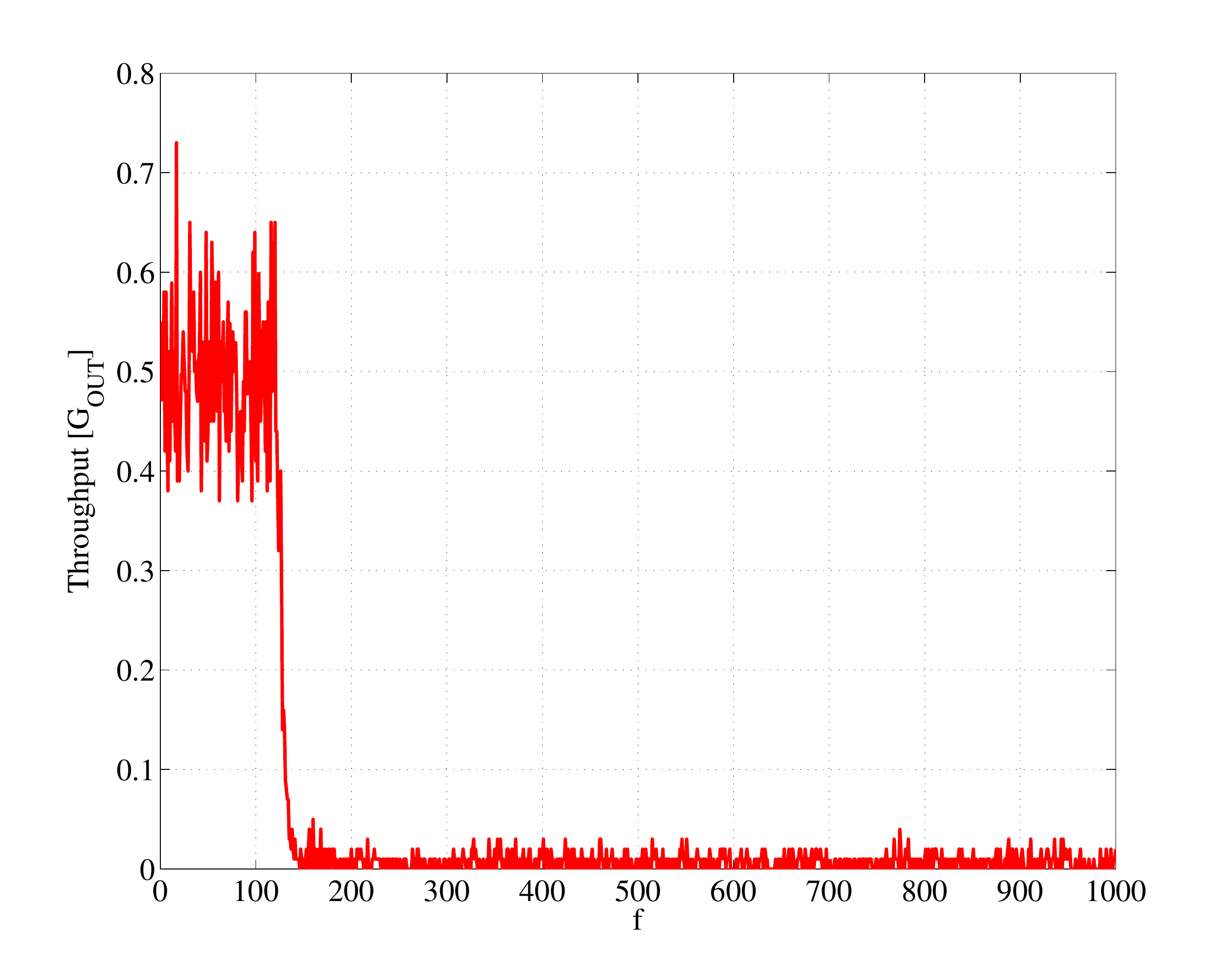}
\caption{Simulation throughput for CRDSA with $N_s=100 \ slots$, $I_{max}=20$, $p_0=0.143$, $p_r=1$, $M=350$ and $f$ indicating the frame sequential number (i.e. the time from the beginning of the communication in frame time units.)}
\label{unstFin_sim}
\end{figure}

Figures~\ref{unstFin} and~\ref{unstInf} show two unstable channels respectively for finite and infinite number of users. Analyzing this two figures for increasing number of backlogged packets, the first equilibrium point is a stable equilibrium point. Therefore the communication will tend to keep around it as for the stable equilibrium point in Figure~\ref{st} and we can refer to it once again as \textit{channel operating point}. However, this is not the only point of equilibrium since more intersections are present. Therefore, due to the abovementioned statistical fluctuations, the number of backlogged users could pass the second intersection and return to the right part of the plane, causing an unbounded increase of the expected number of backlogged users in the case of infinite $M$ (Figure~\ref{unstInf}) or an increase till a new intersection point is reached in the case of finite $M$ (Figure~\ref{unstFin}). In the latter case, this third intersection point is another stable equilibrium point known as \textit{channel saturation point}, so called because it is a condition in which almost any user is in B state and $G_{OUT}$ approaches zero. In the former case of infinite $M$, $N_B$ will increase indefinitely and we can say that a \textit{channel saturation point} is present for $N_B\rightarrow\infty$. The effect of this saturation is that the related throughput falls off to a value close to zero as illustrated in Figure~\ref{unstFin_sim} for the case of unstable channel with finite $M$.
Finally Figure~\ref{overL} shows the case of an \textit{overloaded} channel. In this case there is only one equilibrium point corresponding to the channel saturation point. Therefore, even though the channel is nominally stable, the point of stability occurs in a non-desired region. For this reason this case is separated and distinguished from what is intended in this work as \textit{stable channel}. 

\section{Design choices and parameter settings}

Given $l$, $N_S$, $I_{max}$, in this section we will extensively discuss the role of $M$, $p_0$, $p_r$ and explain how the communication state changes when changing them. First of all this is important in order to understand the relations to be considered in the design phase. Moreover, the same discussion is introductory to understand how control policies can help to obtain better performance and a stable channel even in case of statistical fluctuations that could yield to instability. 

\begin{figure}[b!]

\subfigure [Changing $p_0$] {\label{changingP0} \includegraphics [ scale = 0.24 ]{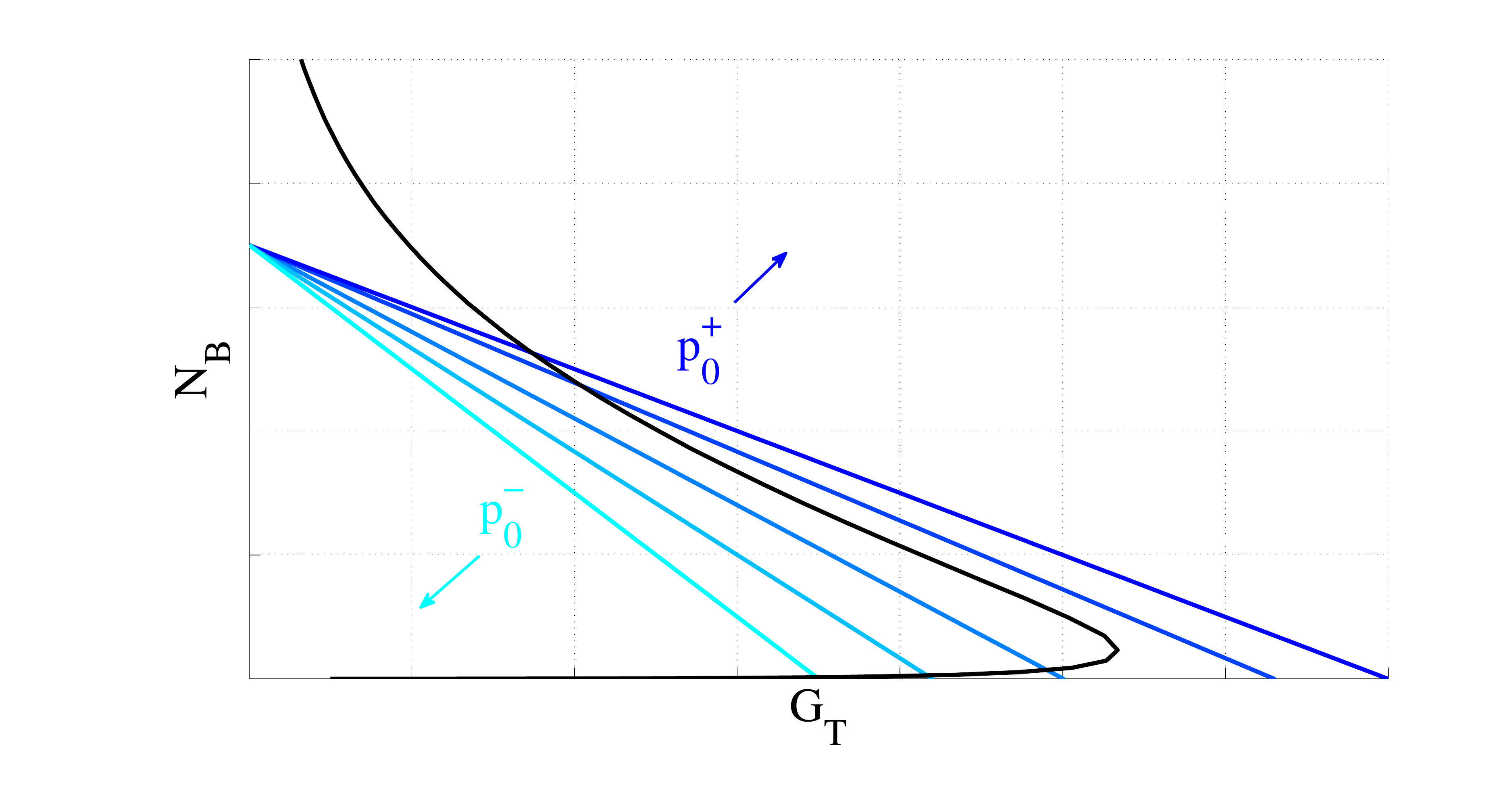}
\label{changingP0}
} \qquad
\subfigure [Changing $M$] {\label{changingM} \includegraphics [ scale = 0.24 ]{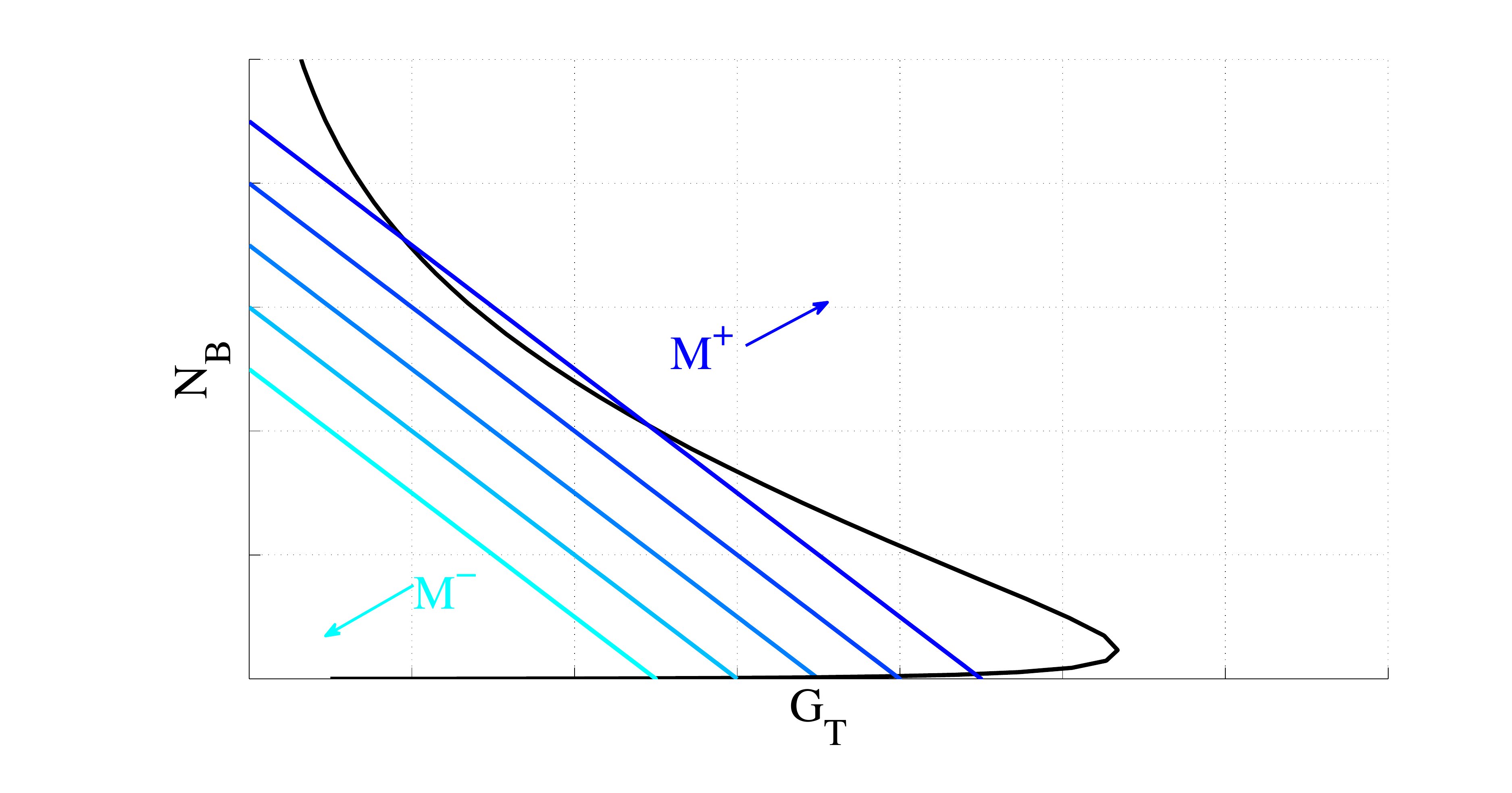}
\label{changingM}
} \qquad
\subfigure [Changing $p_r$] {\label{changingPr} \includegraphics [ scale = 0.24 ]{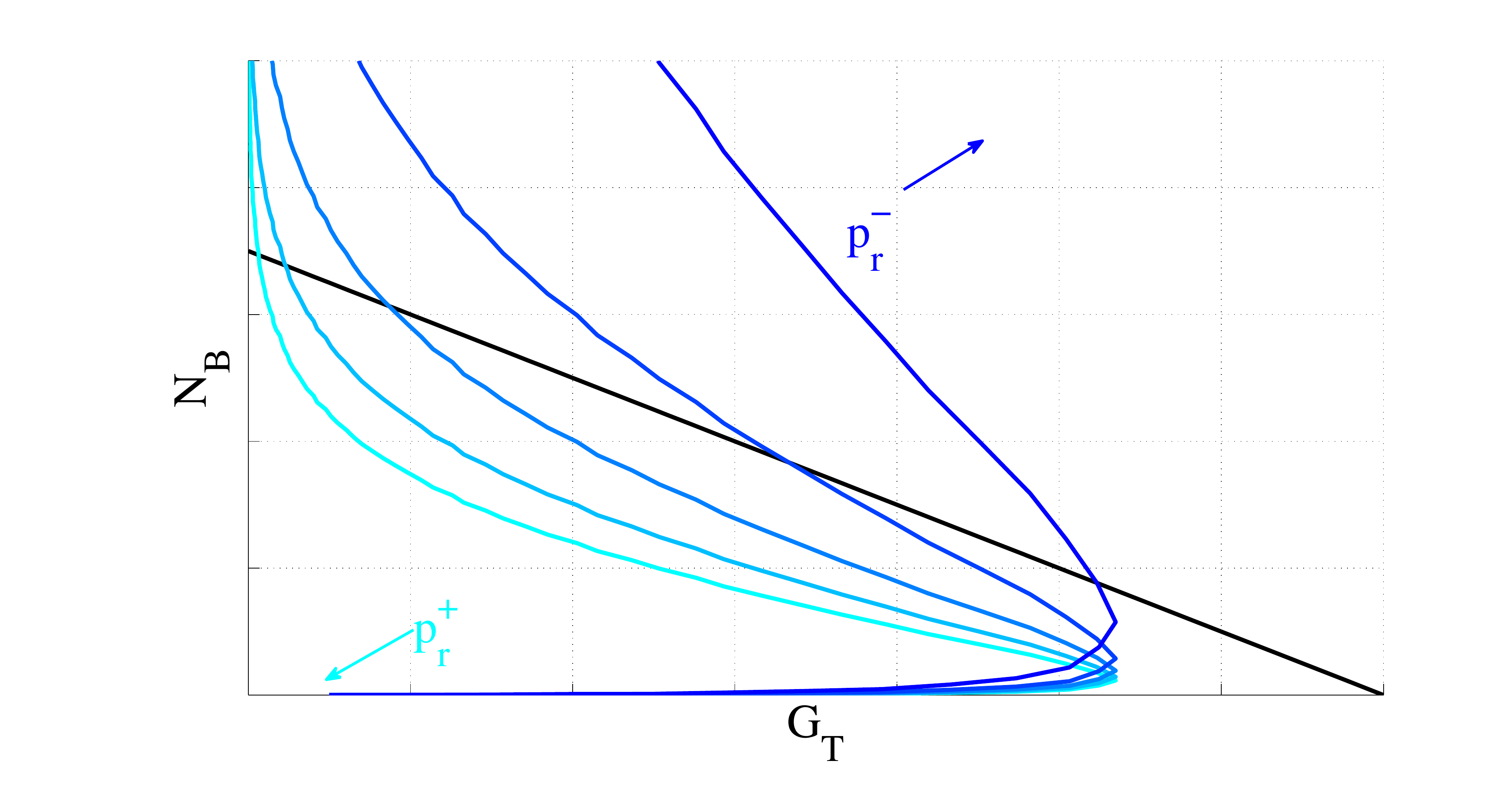}
\label{changingPr}
} \qquad
\caption{Graphical representation of the changes in the ($G_T$,$N_B$) plane when increasing or decreasing one of the parameters' values.}
\label{All_changes}
\end{figure}

Consider Figure~\ref{All_changes}. The first thing we can notice is that $M$ and $p_0$ only influence the channel load line while changing $p_r$ corresponds to a change of the equilibrium contour. In Figure~\ref{changingP0} we can see that $p_0$ influences the slope of the channel load line that becomes steeper when increasing $p_0$. This is intuitive to understand since for fixed $M$, if $p_0^A<p_0^B$ then $G_T^A<G_T^B$. Figure~\ref{changingM} shows that $M$ shifts the channel load line up and down if the population size is respectively increasing or decreasing. From Equation~\ref{LL1} we can notice that the value M is nothing else that what is known in literature as $q$, that is the intersection of the line with the y-axis. Finally $p_r$ determines a shift upward of the channel load line for smaller values of $p_r$. As we can see in Figure~\ref{changingPr} if the value $p_r$ is sufficiently small, it is possible to stabilize a channel initially unstable. This represents the ground base for one of the control policies shown in the next section.

\section{Control Limit Policies}
Previous sections highlighted that statistical variations can yield to instability even in case of initially well-performing channels (see Figure~\ref{unstFin_sim}). To render a channel of that type stable 2 straightforward solutions are available: the first is to use a smaller value for the retransmission probability giving then rise to a larger backlogged population for the same throughput value; the second is to allow a smaller user population size $M$ thus resulting in a waste of capacity. A third solution, that is the one analyzed in this paper, is the use of control limit policies to control unstable channels by applying the countermeasures above in a dynamic manner. In this section we analyze control procedures of the limit type able to ensure stability. This procedures recall those used in \cite{dyn_stab} for SA. In particular two simple yet effective dynamic control procedures are considered: the input control procedure (ICP) and the retransmission control procedure (RCP). These two control procedures are based upon a subclass of policies known as control limit policies, in which the space of the policies is generally composed of two actions and a critical state that determines the switch between them, known as control limit. In this case the control limit is a critical threshold for a certain number of backlogged users $\hat{N}_B$.

\subsection{Input Control Procedure} 
This control procedure deals with new packets to transmit. In particular, two possible actions are possible: accept (action $a$) or deny (action $d$) and the switch between them is determined by the threshold $\hat{N}_B$ as previously mentioned.

\subsection{Retransmission Control Procedure}
As the name says, the retransmission control procedure deals with packets to retransmit and in particular with their retransmission probability. In particular two different retransmission probabilities $p_r$ and $p_c$ are defined that represent respectively the action taken in normal retransmission state (action $r$) and in critical state (action $c$). From the definition above it is straightforward that it must be $p_r>p_c$. The switch between these two modes is determined by the threshold $\hat{N}_B$.\\

\begin{figure}[tbh!]
\subfigure [ICP] {\label{ICP} \includegraphics [ scale = 0.24 ]{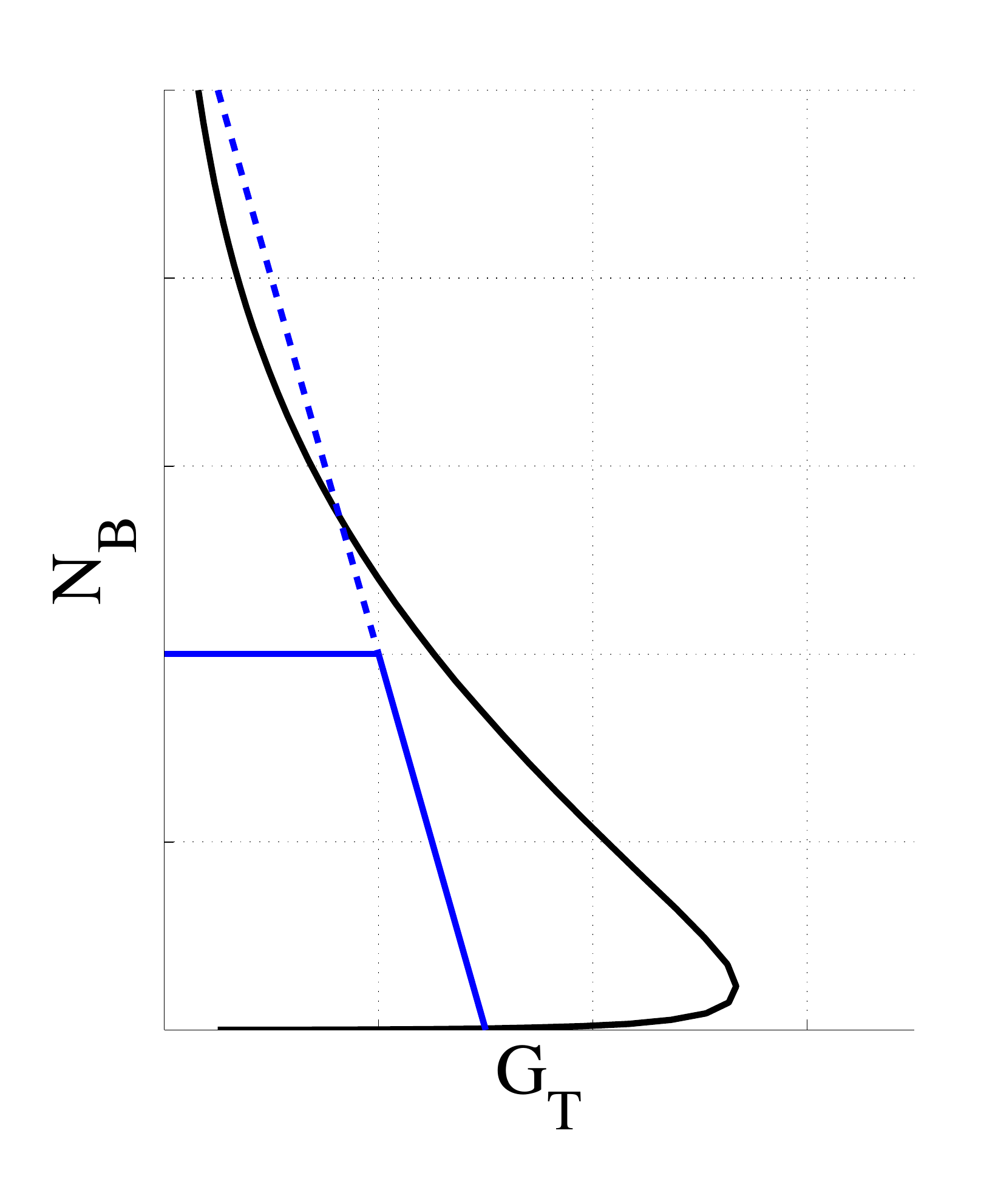}
\label{ICP}
} \qquad
\subfigure [RCP] {\label{RCP} \includegraphics [ scale = 0.24 ]{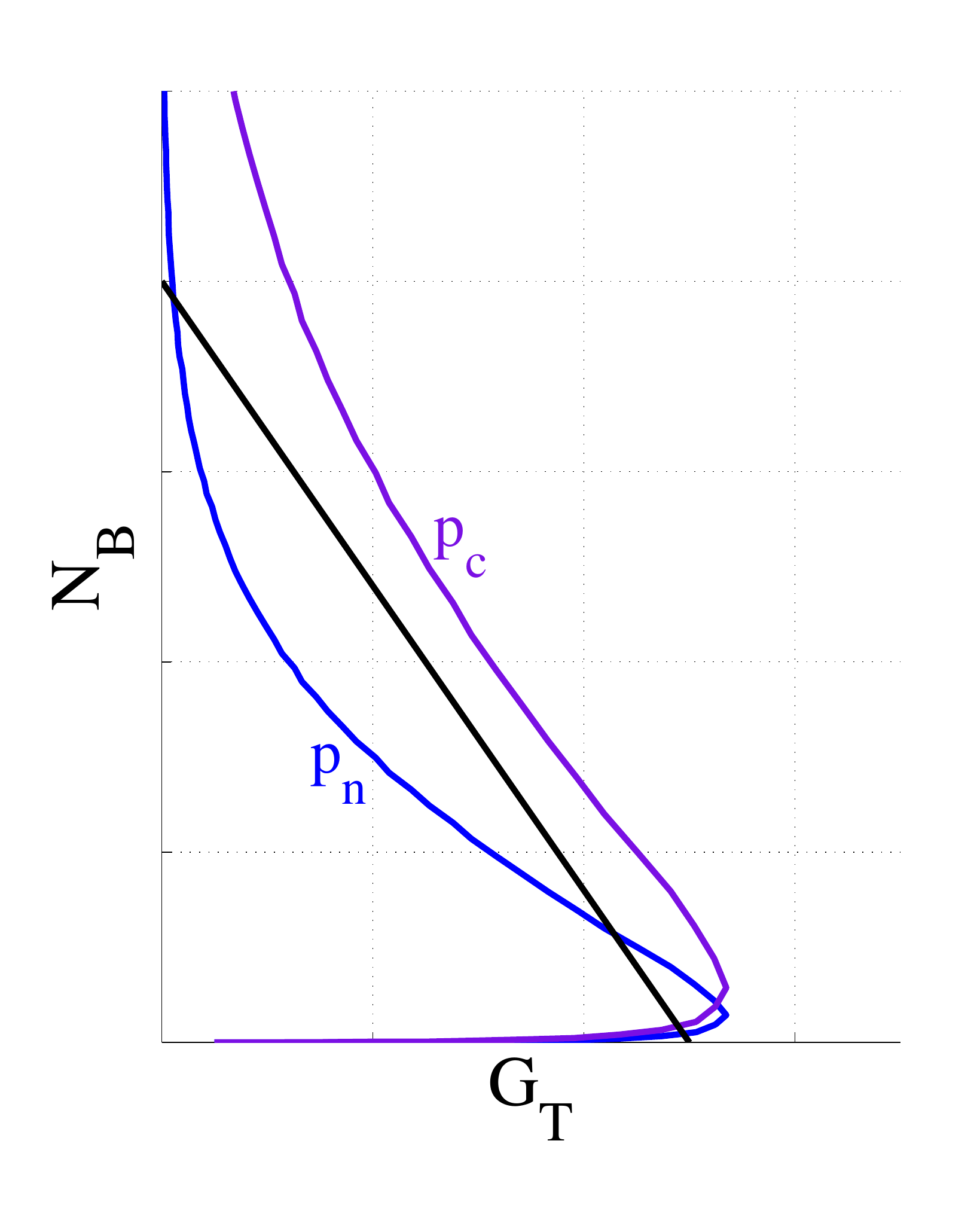}
\label{RCP}
} \qquad
\caption{Control limit policy examples.}
\label{ICPRCP}
\end{figure}

Figure~\ref{ICPRCP} shows two examples corresponding respectively to channels using ICP and RCP. As we can see both policies accomplish the same task of ensuring channel stability. However, while ICP controls the access of thinking users, RCP controls the access of backlogged users.

\section{Simulation Results}

In this section we consider some cases of unstable channels that call for a control policy able to prevent them from saturation. In all the cases analyzed $M$ and $p_0$ are fixed (stationary) and we consider $p_r=1$ and $p_c$ small enough to ensure a stable channel. All the scenarios have been simulated for $10^6$ frames. The three scenarios considered are the followings:
\begin{enumerate}
\item CRDSA with 3 replicas, $M=300$, $p_0=0.2$, $p_c=0.39$ resulting in $N_B^U=25$ when $p_r=1$;
\item CRDSA with 3 replicas, $M=200$, $p_0=0.34$, $p_c=0.5$ resulting in $N_B^U=12$ when $p_r=1$;
\item CRDSA with 2 replicas, $M=220$, $p_0=0.25$, $p_c=0.8$ resulting in $N_B^U=39$ when $p_r=1$.
\end{enumerate}

\begin{figure}[tbh!]
\centering
\includegraphics [width=9 cm] {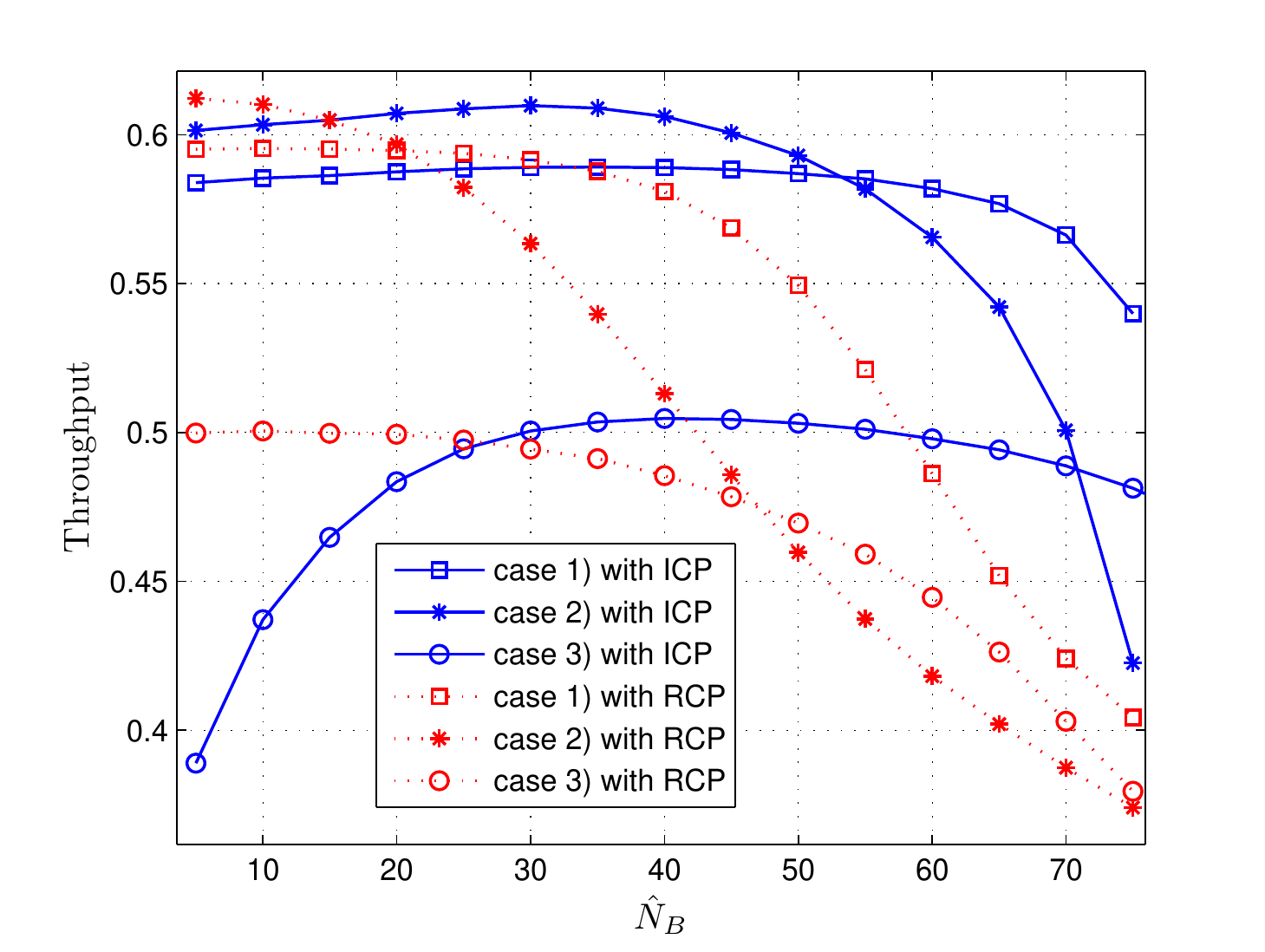}
\caption{Average throughput versus $\hat{N}_B$}
\label{TOTAL_thrp}
\end{figure}

\begin{figure}[tbh!]
\centering
\includegraphics [width=9 cm] {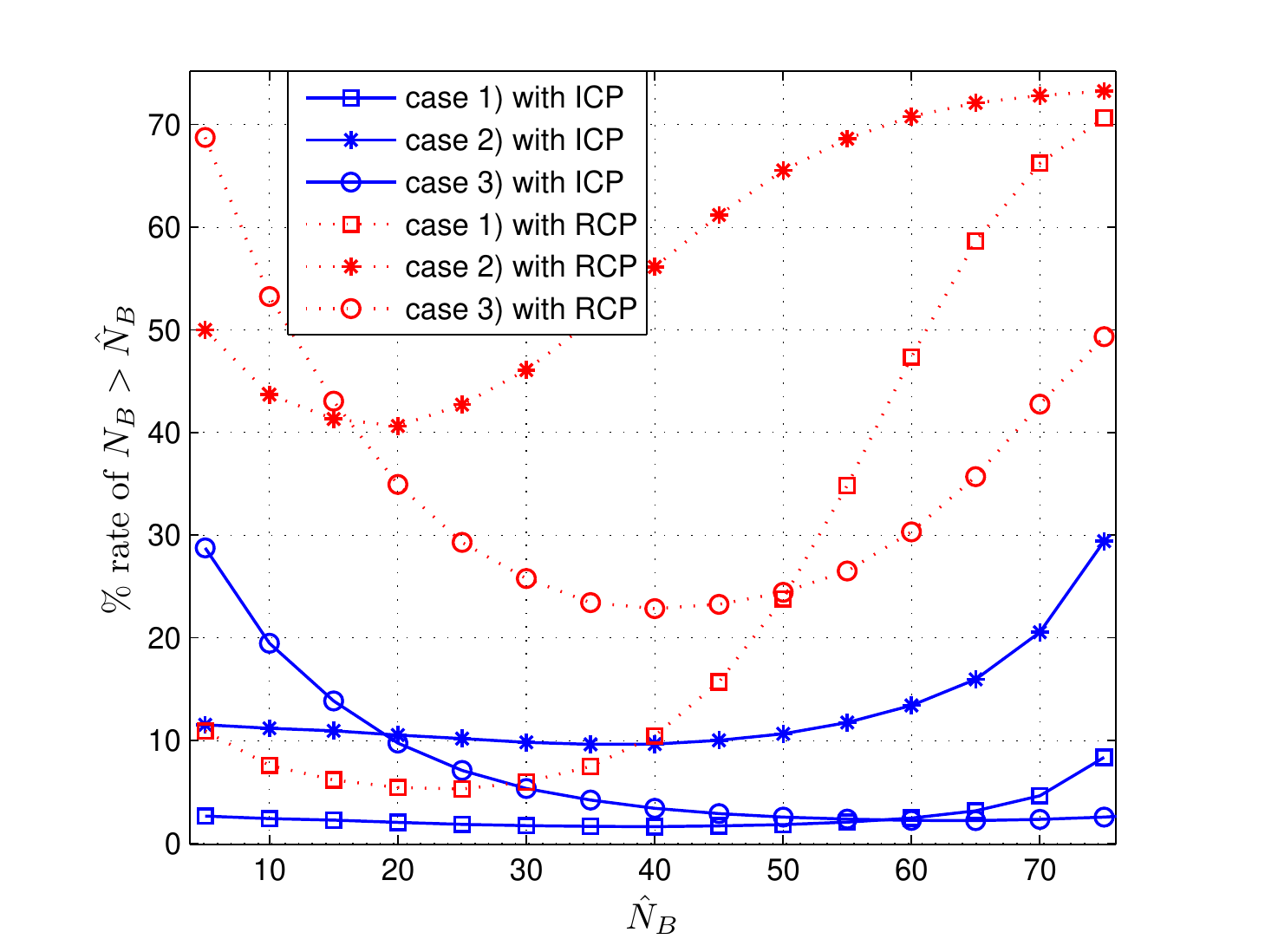}
\caption{Percent rate of application of the limit control policy versus $\hat{N}_B$}
\label{TOTAL_control}
\end{figure}

\begin{figure}[tbh!]
\centering
\includegraphics [width=9 cm] {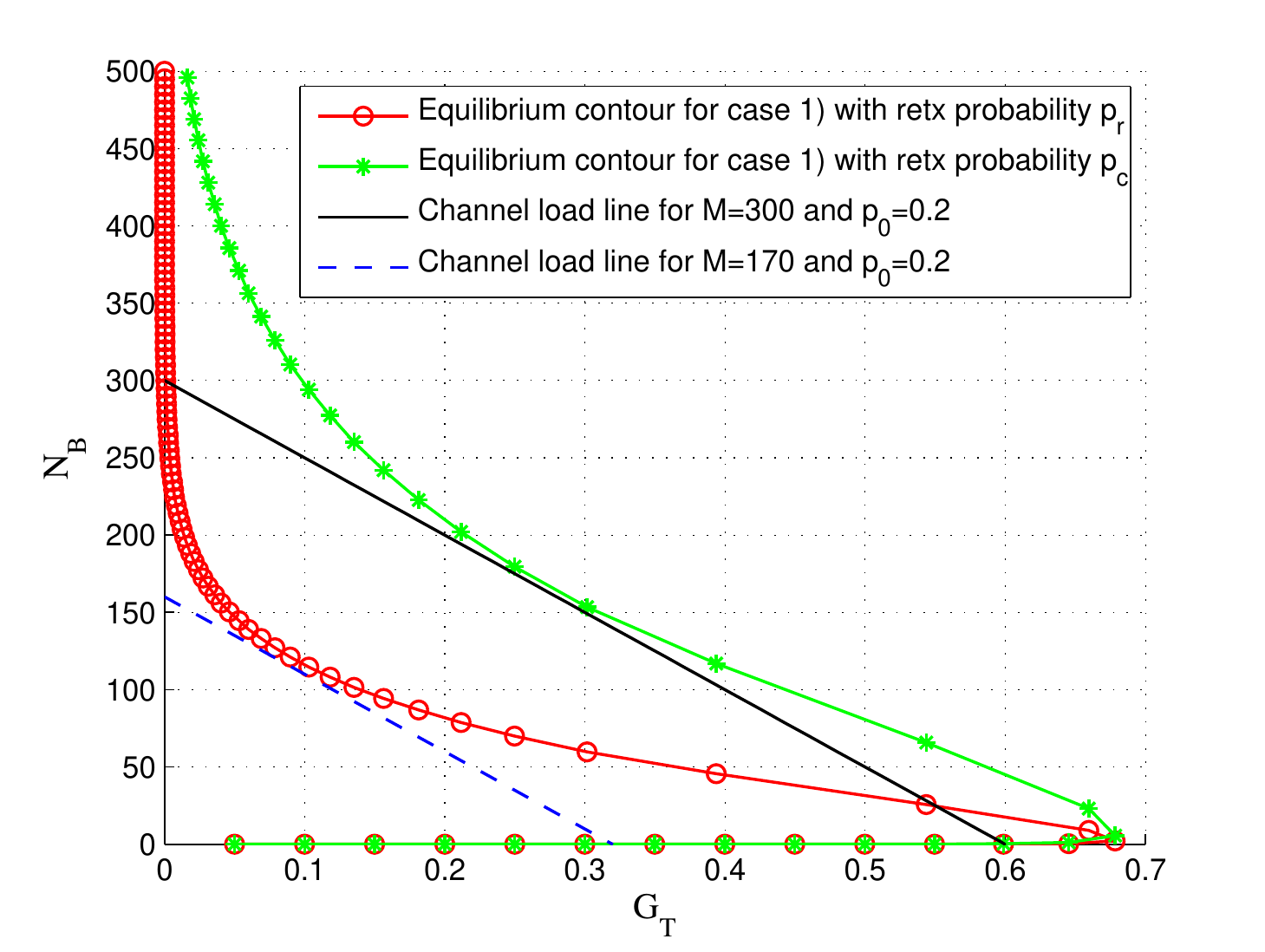}
\caption{Equilibrium contours for CRDSA 3 and channel load lines}
\label{contour}
\end{figure}

Figure~\ref{TOTAL_thrp} illustrates the results in terms of throughput both for ICP and RCP for the study cases enumerated above versus $\hat{N}_B$. Figure~\ref{TOTAL_control} shows the related percentage of frames in which the channel has $N_B>\hat{N}_B$ so that the critical action of the control limit policy is applied ($d$ or $c$ depending on the case). First of all we can notice that both for ICP and for RCP, the range of flatness is wider when operating far from the throughput peak. This is really important for the following reason. Imagine a real application scenario with large delay (as it is the case in satellite communications). Differently from the case analyzed in this paper, when the $\hat{N}_B$ threshold is crossed the control limit policy needs a certain propagation time before it is applied. As a result the effective threshold is not $\hat{N}_B$ but some value $\hat{N}_B+\Delta$, where $\Delta$ depends on the drift of the number of backlogged users for the particular case and on the propagation delay. Therefore the wider is the flat region, the less the performance optimality of the controlled channel will be influenced by the propagation delay. The reason for this difference of flatness can be detected from Figure~\ref{TOTAL_control} where it can be noticed that in general terms  the number of time spent in critical state is greater when closer to the throughput peak. This comes from the fact that the operating point of stability and the point of instability $N_B^U$ are generally closer when approaching the throughput peak while the PLR is greater thus determining are more probable drift beyond $N_B^U$.

Let us now consider the difference between the case with 2 and 3 replicas. As we can see, in the case of 2 replicas a degradation of the throughput and a high percentage of time spent in the critical state is also found for a small $\hat{N}_B$ value. In general terms, when a control limit policy is applied it does not make sense to choose $\hat{N}_B<N_B^U$, therefore this result does not seem to be of interest for our analysis. However let us consider once again the case with large propagation delay. If $N_B>\hat{N}_B$ the control limit policy will switch to critical state until $N_B<\hat{N}_B$ is verified. But as previously mentioned users do not immediately switch state due to the propagation delay, so that in this case the critical policy is applied for a certain period also when $N_B$ is smaller than $\hat{N}_B$ and this could result in suboptimal performance especially in the case of ICP as shown in Figure~\ref{TOTAL_thrp}.

Last but not least let us discuss the results when instead of dynamic control limit policy, the channel is designed by statically limiting the number of allowed users or by decreasing the retransmission probability so that the channel is always stable. Figure~\ref{contour} shows the case of CRDSA with 3 replicas, $M=300$, $p_0=0.2$, $p_c=0.39$. In the first case, the number of allowed users must be decreased to approximately $M=170$ so that the resulting average throughput is around $0.32$ . The second case consists in decreasing the retransmission probability (as done in Figure~\ref{contour} to 0.39). Also in this second case the resulting throughput is diminished, although most of the times this decrement is of small entity. However another effect is that the average packet delay is increased. Therefore depending on the application also this second solution could be problematic.

\section{Conclusions and Future Work}

In this paper a model for computation of the channel stability when using Contention Resolution Diversity Slotted Aloha as Random Access mode in DVB-RCS2 has been outlined. After analyzing the influence of some crucial parameters such as the transmission and retransmission probability and the population size, the paper introduces the concept of control limit policies for such a technique and outlines empirical yet effective procedures for congestion control in case of channels that present the possibility of drifting away to a zone of saturation with low throughput. Finally, simulation results demonstrate how dynamic control limit procedures can help avoiding congestion while allowing the channel to behave in a optimal manner despite the possibility of instability and also in presence of large propagation delay. This work has mainly concentrated on the throughput resulting from the application of this channel procedures. However this represents only half of the story since also the related packet delay associated to successful packets is of utmost interest. For this reason we are currently preparing an extended version of this work where also this second aspect is analyzed together with other important points that have been omitted here for the sake of concision.

\end{document}